# A search for optical laser emission from Alpha Centauri AB

G. W. Marcy ⭐

*Center for Space Laser Awareness, 3388 Petaluma Hill Road, Santa Rosa, CA 95404, USA*

## ABSTRACT

A search for laser light from the directions of Alpha Centauri A and B was performed by examining 15 362 optical, high-resolution spectra obtained between 2004 and 2018. None of the spectra exhibit laser emission lines. The threshold was 10 per cent of the continuum intensity of the spectra of both stars at all wavelengths between 3850 and 6900 Å. This search would have revealed optical laser light from the directions of Alpha Cen B if the laser had a power of at least 1.4–5.4 MW (depending on wavelength) and was positioned within the 1 arcsec field of view (projecting to 1.3 au), for a benchmark 10-m laser launcher. For Alpha Cen A, the laser power must be three times greater for detection. Lasers of smaller aperture would also have been detected but would require more power. Considering all optical surveys, a growing desert is emerging in the search for extraterrestrial technology.

## 1. INTRODUCTION

Speculative models of the Milky Way Galaxy include communication networks composed of transmitters, receivers, and nodes stationed near and between stars (e.g. Bracewell 1973; Freitas 1980; Forgan 2014; Maccone 2014, 2022; Gertz 2021; Hippke 2021; Gertz & Marcy 2022). Network communication may be performed by radio waves (e.g. Enriquez et al. 2017; Price et al. 2020; Margot et al. 2021; Gajjar et al. 2022). Alternatively, lasers may carry information to optimize privacy, bandwidth, and payload (Schwartz & Townes 1961; Zuckerman 1985; Hippke 2018, 2021; Gertz & Marcy 2022), and they may employ ultraviolet, optical, or infrared wavelengths.

A search for extremely rare signals must minimize false positives that would overwhelm the follow-up effort. Lasers emit light at wavelengths that are arbitrary and not emitted by astrophysical objects, able to catch the eye to warrant laborious follow-up work. Modern lasers provide more than enough power, $\sim 10^{15}$ W, for interstellar transmission and yet emit within a narrow range of wavelengths (e.g. Su et al. 2014; Naderi et al. 2016; Lureau et al. 2020; Wang et al. 2020; Liu et al. 2021; Antonio, private communication). Thus, even current technology makes lasers suitable as interstellar couriers, and their arbitrary wavelengths make them nearly unambiguous in the search for extraterrestrial technology (SETI).

Searches for extraterrestrial optical and near-infrared lasers have been carried out using 1-m telescopes to detect pulses of light (Wright et al. 2001; Howard et al. 2007; Maire et al. 2019). Searches for lasers, both pulsed and continuous, have also been performed using high-resolution spectra of more than 7000 stars having all masses, ages, and chemical compositions, including all spectral types OBAFGKM (Reines and Marcy 2002; Tellis & Marcy 2015, 2017; Marcy 2021; Tellis et al., in preparation). No detected lasers emerged, nor even compelling candidates. In addition, a 10 × 14 deg² region towards the Galactic Centre was recently searched for optical laser emission (Marcy, Tellis & Wishnow 2022). All of these optical SETI spectroscopic searches involved acquisition and analysis of spectra in the wavelength range $\lambda \sim$ 3600–9500 Å to detect narrowband emission. The required laser power for detection is typically 1–10 MW, for stars residing between 10 and 100 pc away, assuming a benchmark laser with a 10 m aperture. No optical lasers have been found.

A new perspective on laser communication is emerging that does not involve 'civilizations' or Drake's '*N*'. Laser communication in the Galaxy may be accomplished by a dense network of communication nodes located near stars, including uplinks from planet surveillance probes. This network consists of interstellar transceivers separated by distances measured in light-years (e.g. Gertz 2018). This network can be optimized by placing repeater stations spaced less than 1light-year apart to reduce the decline of laser intensity with distance squared (Gertz & Marcy 2022). A series of repeater stations between the Solar system and Alpha Centauri AB would be detectable by the spillover of laser light at each repeater station because the beam footprint would be much wider than the receivers (Gertz & Marcy 2022). However, none of the past spectra analysed for lasers included Alpha Cen (Tellis & Marcy 2015, 2017). This paper describes a search for such optical laser emission in 15 362 optical spectra in the direction of the nearest stars, Alpha Centauri A and B.

## 2. OBSERVATIONAL METHOD

We retrieved 15 056 spectra of Alpha Cen B and 306 spectra of Alpha Cen A, all obtained with the HARPS spectrometer between 2004 September 12 and 2018 May 2, and kindly made available on the European Southern Observatory (ESO) public data archive (archive.eso.org). The HARPS high-resolution echelle spectrometer resides in a vacuum chamber and is fed starlight via an optical fibre from the ESO 3.6-m telescope at the La Silla Observatory (Mayor et al. 2003), providing outstanding stability of both the wavelength

⭐ E-mail: geoff.bnb@gmail.com

scale and point spread function (PSF) of the spectra. We obtained the archived, reduced, one-dimensional (1D) extracted spectra in units of photons collected in each rebinned 'pixel' that spans 0.010 Å, the standard format.

The wavelength scale is in the frame of the Solar system barycentre and it is accurate to within 0.1 ms$^{-1}$ (European Southern Observatory, HARPS User Manual, 2011). This wavelength scale has a constant offset in velocity from the wavelengths that would be seen at the centre of mass of the Alpha Cen system. Any emitted wavelength that is constant in time in the frame of the centre of mass of the Alpha Centauri system will have a constant wavelength in all of these spectra, albeit shifted due to the Doppler effect between the Sun and the Alpha Cen system. The spectra have a spectral resolution of $\lambda/\Delta\lambda = 115\,000$ and span wavelengths $\lambda = 3781$–6913 Å. All spectra were obtained with exposure times of 4–20 s and reduced by the standard HARPS pipeline. There are typically $6 \times 10^4$ photons per pixel in the blue and green regions and $1.1 \times 10^5$ photons per pixel in the yellow and red regions, yielding a photon-limited signal-to-noise ratio of more than 100, except shortward of 4500 Å where declining throughput causes a declining photon-limited signal-to-noise ratio.

Fig. 1 shows a typical spectrum of Alpha Centauri B on a flux scale normalized to the continuum. The normalization was accomplished with a spline fit to the median of the continuum flux at 40 wavelength regions across the spectrum. The high spectral resolution ($R = 115\,000$) reveals thousands of atomic and molecular absorption lines (Fig. 1), typical for stars of roughly solar mass (within 30 per cent) and having ages 1–10 Gyr (e.g. normal stars of spectral types G2V and K1V). Such stars exhibit no intrinsic emission lines in their spectra. Any technological laser emission line would stand out as non-natural.

We checked the claimed spectral resolution for HARPS of $\lambda/\Delta\lambda = 115\,000$ by examining spectral lines that are intrinsically narrower than the nominal instrumental profile, notably night-sky emission lines [OI] 5577 and [OI] 6300 and telluric O$_2$ absorption lines in spectra of Proxima Centauri (Marcy 2021). Those diagnostic lines displayed a full width at half-maximum (FWHM) of 6 pixels, corresponding to 0.060 Å, consistent with the claimed spectral resolution. Thus, any viable candidate laser emission must exhibit an FWHM $> \lambda/115000$ that results from the convolution of the instrumental profile and the laser line profile. We did not measure the shape of the instrumental profile, nor its constancy in wavelength and time. The quality optics of HARPS and the efficient scrambling by the input fibre indicate that the instrumental profile varies with wavelength and time by less than 10 per cent, as shown in figs 2 and 5 in Marcy (2021).

The entrance of the optical fibre that feeds the HARPS spectrometer has a field of view of 1.0 arcsec, and it is pointed directly at either Alpha Cen A or B for these observations. In 2015 May, the fibres were replaced with new fibres having an octagonal cross-section of nearly identical field of view. As Alpha Centauri is 1.34 pc from the Earth, the 1.0 arcsec field of view expands to a circular footprint 1.34 au in diameter there and accepts light emitted from within the cone anywhere along the line of sight. The two stars are observed in separate exposures, thereby sampling separate conical fields of view. Each individual spectrum can reveal technological light that is emitted from within the 1.0 arcsec cone and directed toward the telescope. Light originating outside that cone obviously cannot be detected. The two fields of view extend into the background behind each star.

During the 13.6 yr of observations, Alpha Cen A and B moved in their orbit around each other. The orbit has an apparent relative semimajor axis of 17.5 arcsec (maximum angular separation), an orbital period of 79.8 yr, and an orbital eccentricity of 0.52. Each of the 15 362 spectra taken during the 13.6 yr samples had a different 1.0 arcsec field of view because of both their orbital motion and the orbital motion of the Earth around the Sun that causes 1.3 arcsec of parallax. Between 2004 and 2018, spanning the observations here, the angular separation between the two stars changed from 11 to 5 arcsec, with each star moving along a curved portion of their mutual orbit. The telescope points toward where each star was 4.37 yr prior, to account for the light traveltime, thereby tracking their orbital motion. Thus, the 15 362 spectra sample light from a region of two curved paths within a region ∼11 arcsec across as the two stars execute their mutual orbit about the centre of mass.

The parallax of 1.34 arcsec due to the Earth's orbital motion around the Sun also plays a role. Alpha Cen resides at ecliptic latitude 43.5 South, implying our vantage point changes annually in a projected oval pattern having a total extent of ∼2.6 arcsec (after 6 months) toward each star. The changing parallax during a year widens (i.e. smears) the ensemble of fields of view superimposed on the binary orbit. In summary, parallax and the binary motion of Alpha Cen A and B cause the 15 362 observations to spectroscopically survey two swaths of space that are each ∼2 arcsec wide and ∼6 arcsec long.

The Earth never resides on the line connecting the Sun to the centre of mass of Alpha Cen AB, but instead is between 0.85 and 1.0 au away from that line. Any light beam launched along the line connecting the centre of mass of Alpha Cen to the Sun, and that is directed within 1 arcsec of the Sun, will miss the Earth.

## 3. SEARCH FOR LASER LINES

A method was created to search for laser emission in these high-resolution spectra of Alpha Cen A and B. A continuum-normalized spectrum was computed using a cubic spline fit to a 100-pixel median-smoothed version of the spectrum. (Thanks is due to Gibor Basri for this 'contf' code, dated 1988). Each spectrum was divided by its median-smoothed spectrum to produce a spectrum that has its pseudo-continuum normalized to 1.0, as shown in Fig. 1.

The detection threshold for laser emission lines was established by examining a few hundred spectra for typical departures above the normalized continuum caused by Poisson noise, flat-field errors, or inaccuracies in the continuum determination, all of which causing rare departures of ∼1 per cent, especially near the dense and strong stellar absorption lines between 3850 and 4300 Å. These departures led to a '10$\sigma$' threshold of 10 per cent above the continuum, as a trade between the lowest possible detection threshold (near 1 per cent above continuum) and 20 per cent that would yield no false positives, even in the choppy near-UV spectrum.

The detection algorithm involved three steps. First, each spectrum was median smoothed with a width of 5 pixels. This has the effect of removing almost all elementary particle hits on the CCD and any other defects that can create fake emission features having a width less than 2 pixels. Any emission lines wider than 3 pixels (FWHM) survive this smoothing. Real emission lines must have an FWHM > 6 pixels, due to the width of the instrumental profile convolved with the intrinsic width of the emission line. Indeed, laser lines have intrinsic widths due to the Heisenberg uncertainty principle, the confined size of the laser chamber, and any engineering and optical imperfections. Thus, laser lines will have a measured FWHM greater than ∼6 pixels, and the 5-pixel median smoothing will suppress the peak height by less than 20 per cent. This sacrifice of sensitivity is valuable in return for eliminating elementary particles that masquerade as emission lines.

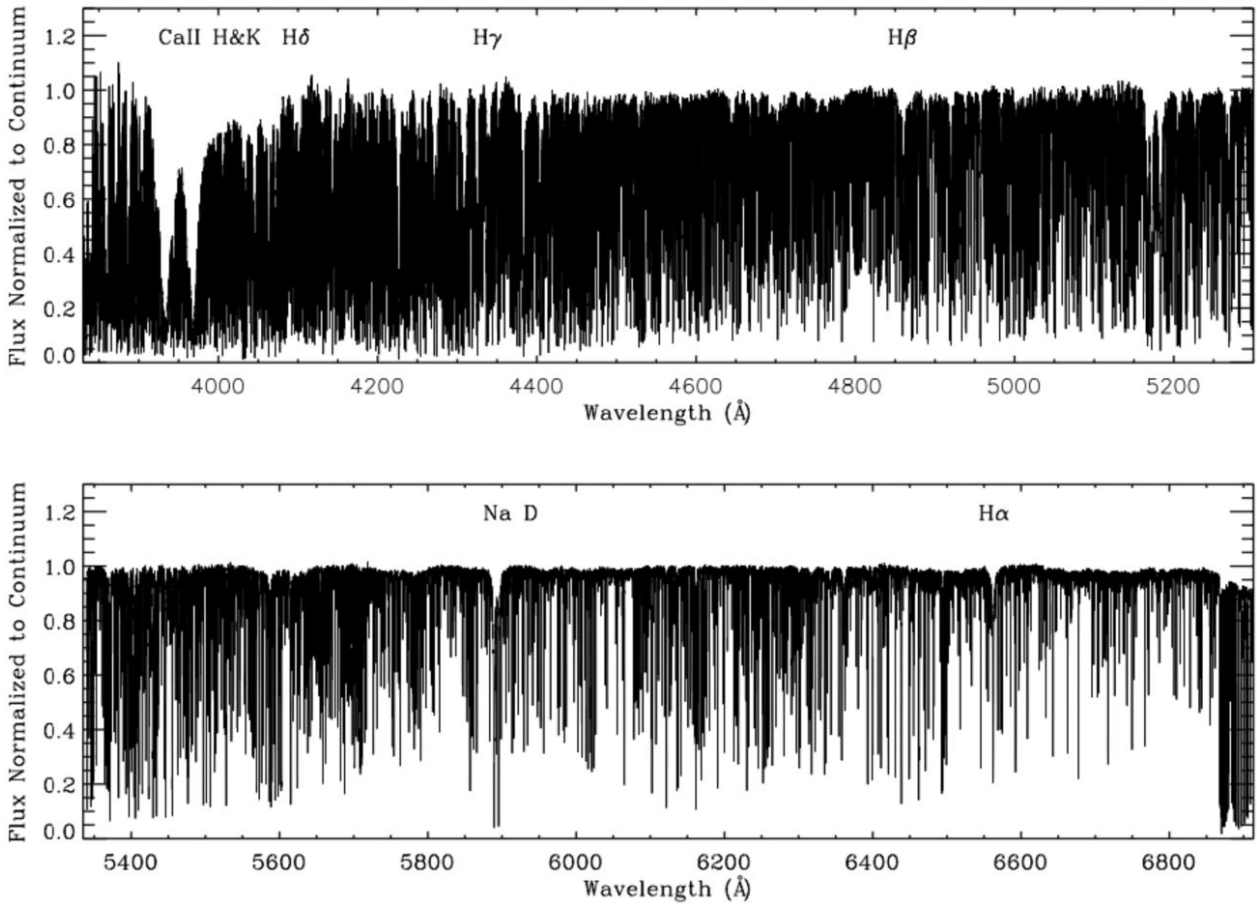

**Figure 1.** Representative optical spectrum of Alpha Centauri B, plotted versus wavelength with the flux normalized to the continuum of the stellar spectrum. Thousands of atomic absorption lines are visible, typical of G and K spectral types. The signal-to-noise ratio is >100 per pixel, except shortward of 4500 Å. No emission lines exist in the spectrum, as is normal for middle-aged stars having roughly solar mass. Emission lines more than 10 per cent of the continuum level (i.e. above 1.1) would be detectable and extraordinary.

Second, code was written to examine all 15 362 spectra for any pixel with a flux 10 per cent above the star's continuum level, i.e. above 1.10 in the normalized spectra. Third, all candidate emission lines were examined by eye. Both narrow and broad emissions up to 5 Å wide will be detected if the peak is 10 per cent above the continuum. Broad emission can occur either by a tight cluster of narrow laser lines smeared by the spectrometer's instrumental profile or by one wide emission line. In summary, the detection algorithm is sensitive to emission widths from 0.06 to 5.0 Å having a peak intensity 10 per cent above the star's spectrum.

This detection algorithm for emission lines was executed on all 15 362 spectra of Alpha Cen AB, yielding eight candidates shown in Fig. 2. Further vetting involves assessing if a candidate emission-line profile is consistent with the instrumental profile, especially the key properties of an FWHM > 6 pixels and asymptotically declining wings (Marcy 2021). Examination by eye of the eight candidates shows that only the emission line shown in the upper right panel of Fig. 2 is consistent with the properties of the instrumental profile. All of the other candidates exhibit profiles narrower than an FWHM of 6 pixels. Those seven rejected candidates were likely caused by incoming elementary particles or gamma rays, or by defects in the CCD detector. We reported more elementary particles in the previous paper (Marcy et al. 2021) about Proxima Centauri, which is 10 mag fainter than the two stars, Alpha Cen AB. High energy particles can deposit enough electrons to be visible in spectra of the faint Proxima Cen but not enough electrons to be visible in Alpha Cen AB that are 10 000 times brighter.

The remaining candidate in the upper right panel of Fig. 2 shows a magnified view of 1 Å of the spectrum centred at 5807.45 Å of Alpha Cen B, taken on 2011 April 29. It constitutes a viable emission line caused by light that actually entered the HARPS spectrometer because the emission consists of ∼9 pixels (0.090 Å) and has an FWHM ∼ 6 pixels, which is consistent with the instrumental profile.

To examine this viable candidate at 5807.45 Å, it is helpful to subtract off the reference stellar spectrum of Alpha Cen B, to leave the transient emission by itself. Eight spectra of Alpha Cen B were chosen at random to average together, constituting the reference stellar spectrum. The eight random spectra are shown in Fig. 3, along with the anomalous emission in the upper left panel. The result of the subtraction is shown in Fig. 4. The upper panel shows the same candidate emission spectrum. The middle panel exhibits the reference spectrum, composed of the average of the eight randomly drawn spectra of Alpha Cen B. The bottom panel shows the result of the subtraction, revealing the apparent emission by itself.

The candidate emission line shown in the bottom panel of Fig. 4 has an FWHM ∼ 6.5 pixels, which is consistent with the width of the instrumental profile. However, the shortward wing (on the left-hand side) declines more quickly than the instrumental profile, with a slope that discontinuously changes from a steep decline to flat, showing no evidence of a smoothly declining wing as the instrumental profile would produce. This steep decline and discontinuity of the shortward wing raise a flag of concern about the emission as actually coming from light that passed through the spectrometer.

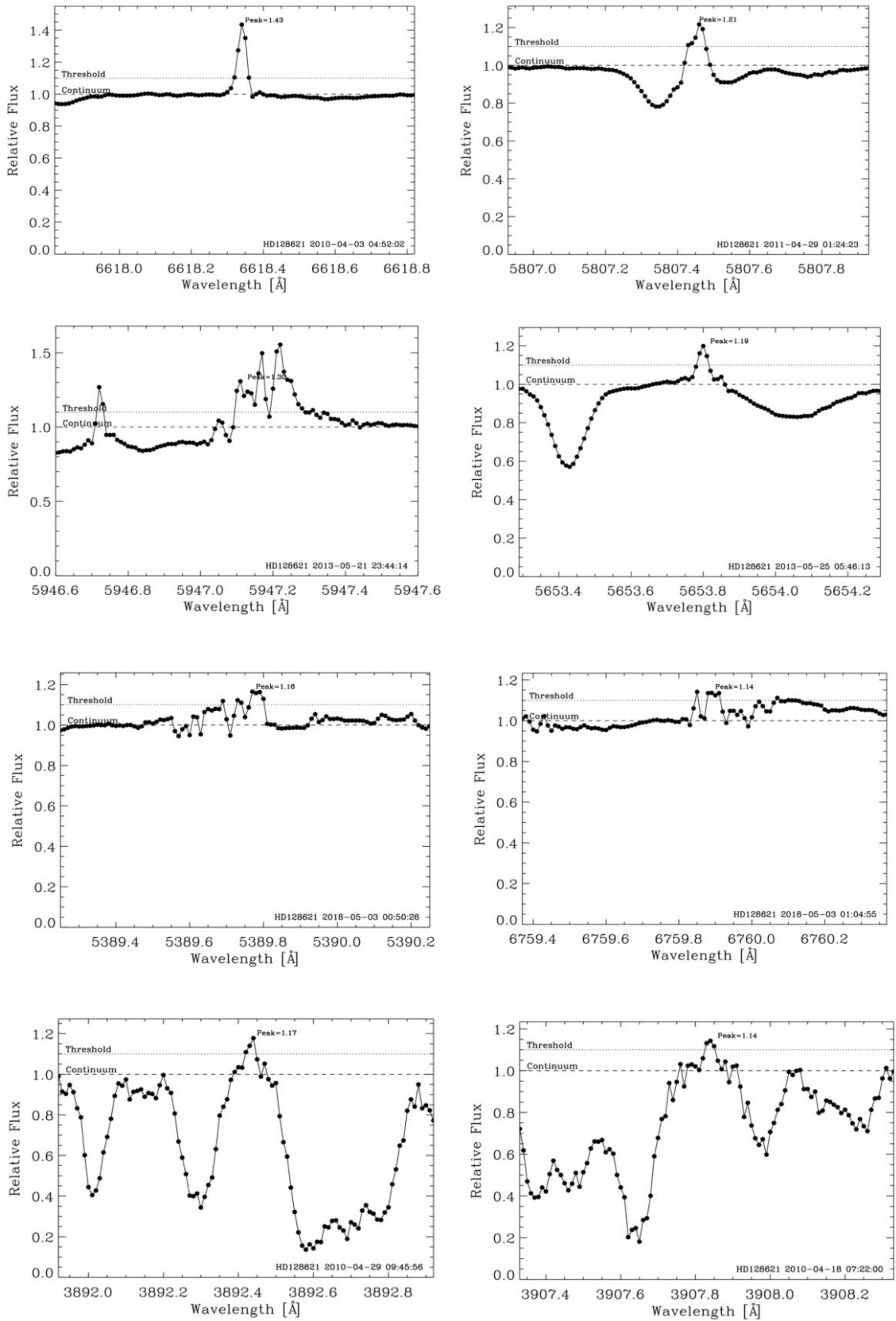

**Figure 2.** The eight candidate emission lines found among the 15 362 spectra of Alpha Cen A and B, from blind execution of the search algorithm. The threshold (labelled) requires that at least one pixel be 10 per cent above the stellar continuum (labelled). A final criterion is that the emission must have an FWHM > 6 pixels, the width of the instrumental profile of the HARPS spectrometer. Only the candidate in the upper right panel at wavelength 5807.4 Å meets that criterion, making it viable.

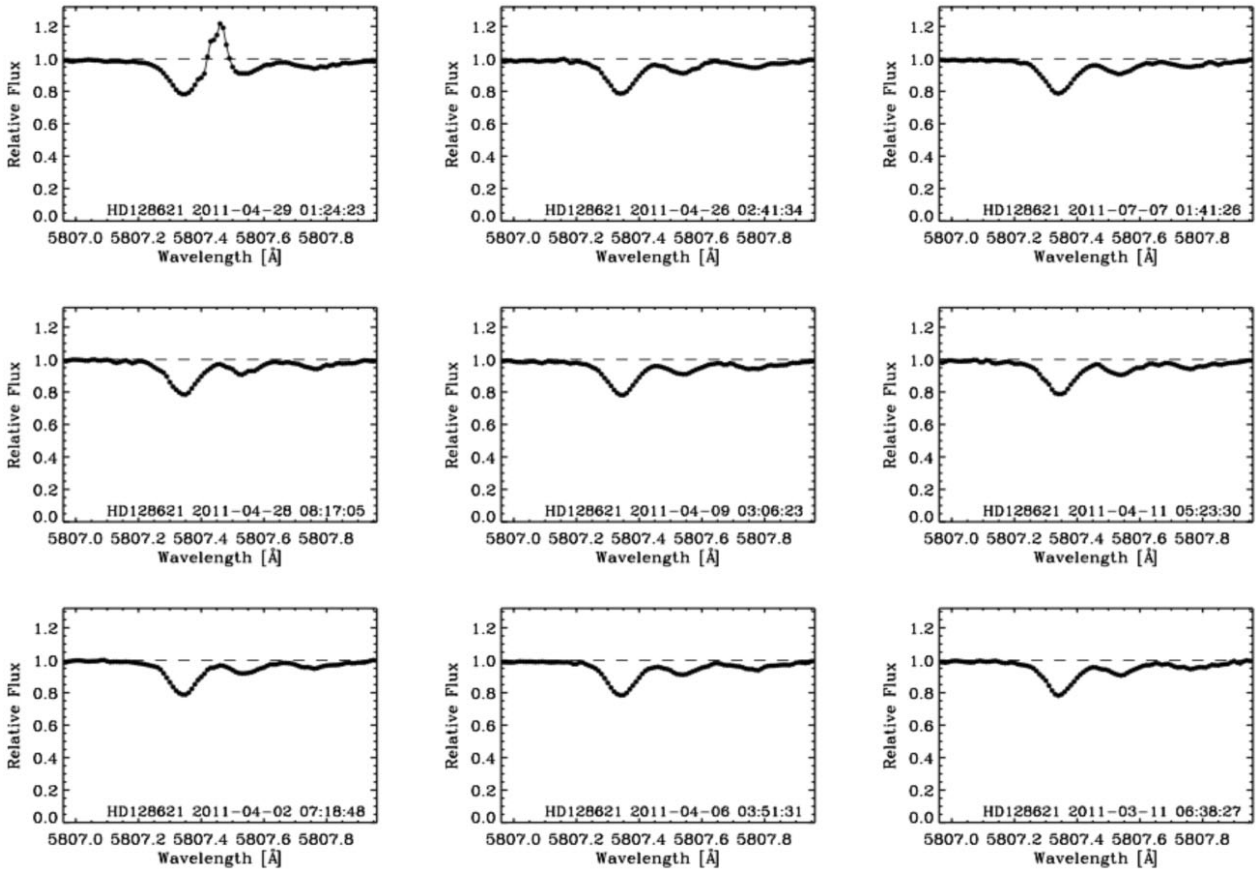

**Figure 3.** The spectrum of Alpha Cen B that has the candidate emission line at 5807.45 Å (upper left panel) compared to eight representative spectra, among the ∼15 000, of Alpha Cen B for comparison. Three weak absorption lines are apparent and constant within the 1 Å region. The emission appears only on 2011 April 29 and is absent in all other spectra.

To understand this candidate emission further, the raw CCD image was examined, generously provided by the ESO public data archive, and is shown in Fig. 5, which shows the emission feature clearly in the raw CCD image. The emission feature is clearly visible as the white region occupying 10 pixels just to the right-hand side of the centre. In this 'lego' plot format, the candidate emission is clearly displaced perpendicular to the horizontal ridge along the stellar continuum spectrum. The candidate emission pixels (white) are displaced one full pixel higher than the horizontal centre (the ridge) of the stellar spectrum. This displacement is in the 'spatial direction'. Specifically, the candidate emission is nearly absent in the bottom two rows of the stellar spectrum. But the emission is bright in the top row of the stellar continuum.

This vertical displacement of one pixel of the candidate emission is inconsistent with the performance of the fibre optic feed of the HARPS spectrum. The fibre scrambles all spatial inhomogeneities in the incoming light within the 1 arcsec field of view. It produces an output beam into the spectrometer that is independent of any spatial structure or non-uniformities in intensity. In particular, the output end of the fibre has a distribution of brightness over its surface that is the same for all neighbouring wavelengths of light, independent of spatial structure at the input end. The output end of the fibre does not necessarily achieve a uniform brightness over its surface. But the distribution of light at the end of the fibre is the same for all nearby wavelengths, for example at 5807.0, 5807.5, and 5808.0 Å. The spectrometer merely images the end of the fibre at each wavelength. Thus, any actual emission line must pass through both the fibre and the spectrometer, and it must end up with no displacement vertically. The apparent displacement of one pixel in Fig. 5 strongly indicates the candidate emission did not result from light that passed through the fibre, and thus was not imaged by the telescope. It can't be caused by light captured by the telescope.

Moreover, Fig. 5 shows the candidate exhibits extremely steep sides that are inconsistent with the Gaussian-like instrumental profile. Thus, two independent properties of the emission are inconsistent with the performance of the fibre and spectrometer. Thus, we must rule out this candidate as an actual emission line coming from the direction of Alpha Cen B.

Alpha Cen A and B exhibit weak emission reversals of chromospheric origin in the cores of the CaII H and K lines, as is common for middle-age G- and K-type stars. Otherwise, no emission lines appear in the spectra. In summary, examination of the 15 362 spectra of Alpha Cen A and B revealed no emission lines having a peak intensity 10 per cent higher than the stellar spectrum at wavelengths 3850–6900 Å. Any source of emission, of either natural or technological origin, located within the 1 arcsec field of view toward either of the stars would have been detected.

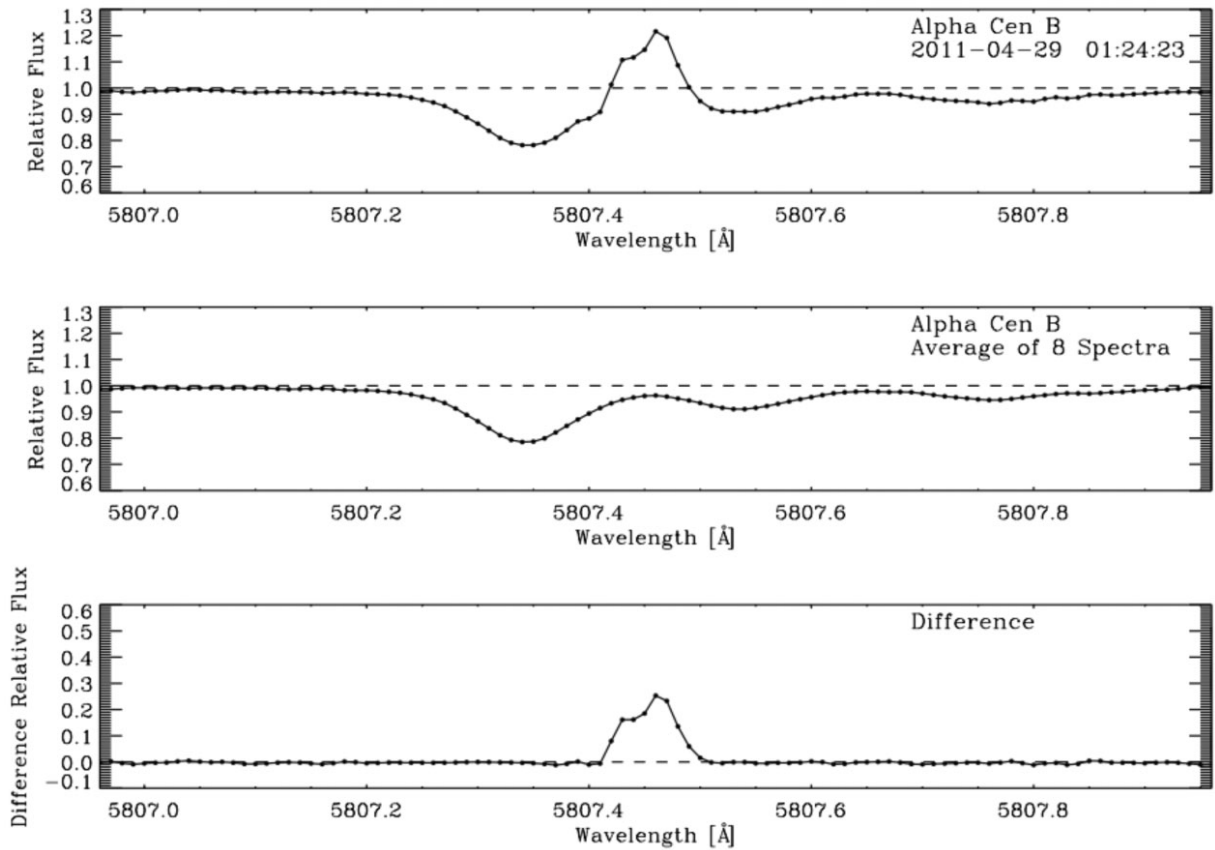

**Figure 4.** Subtracting the stellar spectrum to reveal the transient emission feature by itself. Top panel: Spectrum of Alpha Cen B showing excess flux well above the continuum (dashed line). Middle panel: The average of eight spectra of Alpha Cen B, randomly chosen. Bottom panel: The difference spectrum, revealing the anomalous emission by itself. The sharp drop and lack of a 'wing' on the shortward side is inconsistent with the expected smooth wings of the nearly Gaussian-shaped instrumental profile of the spectrometer, indicating that the apparent emission did not actually pass through the spectrometer.

## 4. LASER POWER THRESHOLDS FOR ALPHA CENTAURI

The detection threshold of 10 per cent above the stellar continuum flux, which varies with wavelength, can be translated to thresholds of laser power. We aim for an accuracy of only ∼30 per cent, as the detailed properties of any laser, such as the wavelength bandwidth and size remain unknown, precluding a more precise calculation. The laser threshold is calculated here for Alpha Cen B, which has a continuum photon flux roughly 1/3 that of Alpha Cen A, and the threshold depends on the flux in the stellar spectrum. Fig. 6 shows the number of photons detected within each pixel in a 10 s exposure of Alpha Cen B for wavelengths from 3850 to 6900 Å. At representative wavelengths of 4000, 5000, and 6000 Å, the continuum (upper envelope) contains ∼1 × $10^4$, ∼4 × $10^4$, and ∼5 × $10^4$ photons per pixel, respectively, to within 10 per cent of accuracy. Any emission lines detected here must have a peak that is 10 per cent higher than those values, i.e. ∼1 × $10^3$, ∼4 × $10^3$, and ∼5 × $10^3$ photons per pixel.

The full laser power includes all wavelengths within the emission profile having a width of at least ∼6 pixels (FWHM), depending linearly on wavelength, due to the instrumental profile. Approximating the unknown profile of the laser emission as a triangle having a base 12 pixels long yields an integrated number of photons given by the number of photons per pixel at the peak multiplied by 6. Applying the peak thresholds listed above, the thresholds within the entire profile are 0.60 × $10^4$, 2.4 × $10^4$, and 3.0 × $10^4$ photons, at 4000, 5000, and 6000 Å, respectively. These values pertain to the 10 s exposure of Fig. 6, yielding rates of photons contained within the profiles of 600, 2400, and 3000 photons $s^{-1}$.

The photon rates above refer to photons detected by the CCD detector. Photons entering the Earth's atmosphere and aimed at the ESO 3.6-m telescope are detected with efficiencies of 2.67 per cent, 5.18 per cent, and 5.07 per cent, including the telescope, losses at the fibre entrance (depending on seeing), the spectrometer, and the CCD (European Southern Observatory 2019). Applying those efficiencies gives thresholds of 2.2 × $10^4$, 4.6 × $10^4$, and 5.9 × $10^4$ photons $s^{-1}$ entering above the atmosphere and aimed at the telescope. For a 3.6-m telescope with a collecting area of 10 $m^2$, the photon rates above the Earth's atmosphere correspond to fluxes of 2.2 × $10^3$, 4.6 × $10^3$, and 5.9 × $10^3$ photons $m^{-2} s^{-1}$, for the three wavelengths, respectively, constituting the threshold photon fluxes.

To produce these threshold photon fluxes, a reference benchmark laser is considered having a diffraction-limited beam opening angle of 1.2 $\lambda/D$ rad (from axis to the first null) and located at the Alpha Cen system. Here, $\lambda$ is the wavelength of light and $D$ is the laser aperture diameter of 10 m. Such a benchmark laser beam intercepts the Earth with a circular footprint having area $A = \pi(1.2\ \lambda\ d/D)^2$, where $d$ is the distance to Alpha Centauri of 1.34 pc. For reference at 5000 Å, the beam footprint at the Solar system has a radius of 0.016 au and an area of 1.9 × $10^{19}$ $m^2$. For scale, this benchmark beam footprint is larger than the cross-section of the Earth but smaller than the inner Solar system.

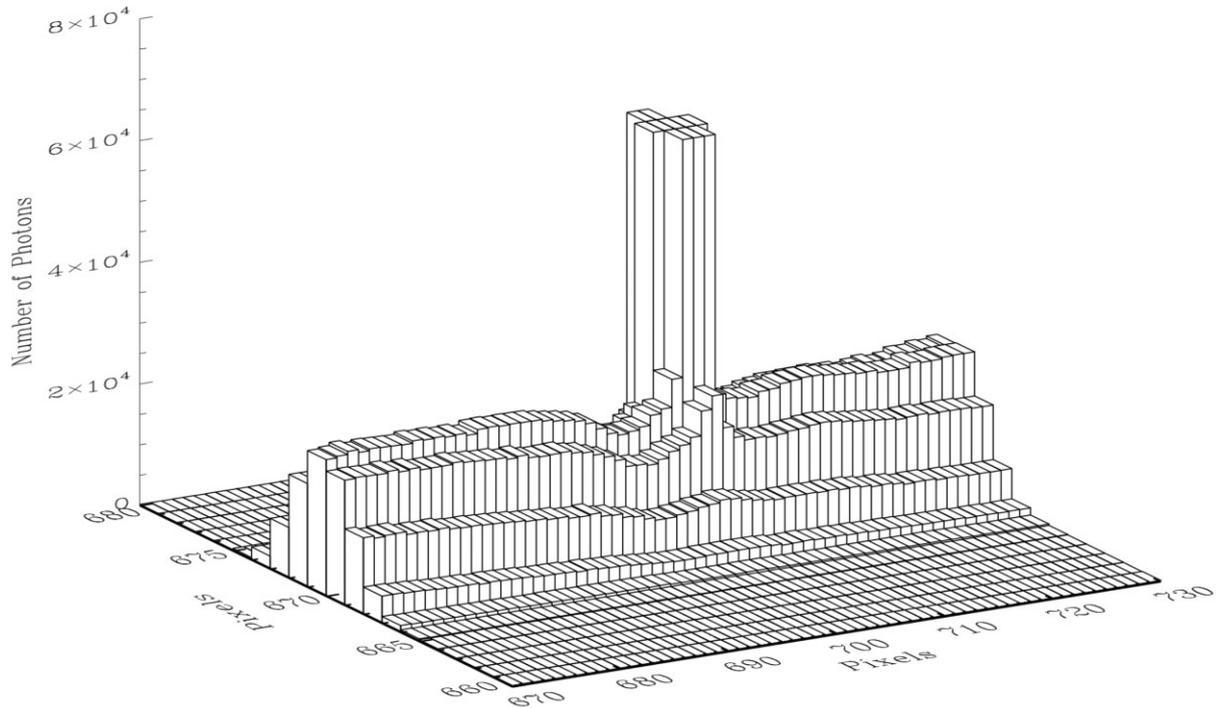

**Figure 5.** The raw image near the candidate monochromatic emission. The 'lego' plot format shows that the candidate is displaced from the ridge along the stellar spectrum, inconsistent with fibre scrambling. Also, the sides of the candidate emission are steep, which is inconsistent with the smooth, Gaussian-like wings of the instrumental profile. This emission cannot be the result of light passing through the spectrometer.

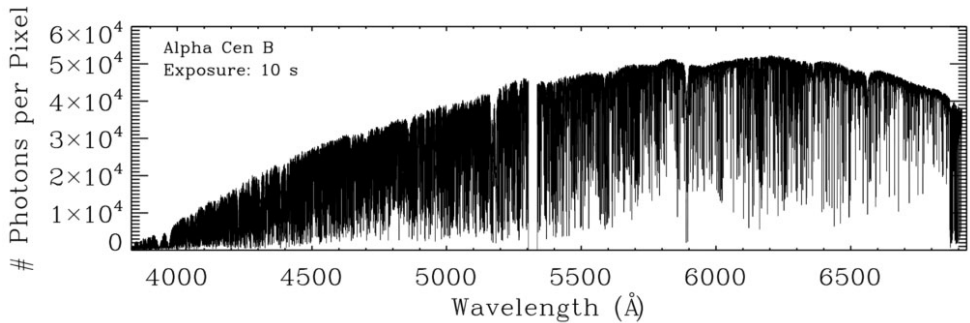

**Figure 6.** Number of photons per pixel in a typical 10 s exposure of Alpha Cen B. The continuum intensity (upper envelope) varies with wavelength. The detection threshold for emission lines is 10 per cent above the continuum level.

Multiplying the threshold photon fluxes entering the atmosphere by the area of the beam footprint at the Solar system yields the total photon rates that must be emitted by the laser, giving $2.7 \times 10^{22}$, $8.9 \times 10^{22}$, and $1.6 \times 10^{23}$ photons s$^{-1}$, respectively, at the three fiducial wavelengths. Multiplying these threshold photon luminosities by the energy per photon gives the laser power required for detection: 1.4, 3.5, and 5.4 MW for assumed wavelengths of 4000, 5000, and 6000 Å, respectively, for Alpha Cen B. For lasers near Alpha Cen A, the required laser power is ∼3 times greater due to its brightness.

For comparison, the required laser power to detect the same benchmark laser at Proxima Centauri is only ∼100 kW (Marcy 2021). This required laser power is considerably lower than that for a benchmark laser at Alpha Cen AB because Proxima has an optical flux $10^{-4}$ times that of Alpha Cen A and B (10 mag fainter at V band), improving the contrast of the laser light against the stellar flux.

For a laser of only 1 m aperture (not 10 m), the required laser power is 100× greater. If a laser with a 10 m aperture were located only 1/100 of the distance to Alpha Cen, as discussed in Gertz & Marcy (2022), the required laser power for detection would be only 133, 348, and 540 W respectively. These low power requirements for detection of lasers stem from the same optimization that supports the close spacing of repeater nodes in the first place (Gertz & Marcy 2022).

## 5. DISCUSSION AND SUMMARY

Examination of 15 362 high-resolution spectra of Alpha Centauri A and B, obtained during the years 2004–2018, revealed no monochromatic emission lines. Lasers having a power of 1.4–5.4 MW at Alpha Centauri B, or three times more for Alpha Centauri A, would have been detected if launched from optics similar to the largest telescopes on Earth. For smaller lasers, the power requirements increase inversely as the square of their size. The laser power requirement of ∼10 MW is modest, suggesting plausible energy sources such as on-board nuclear or recharging with photovoltaics. For repeater transceivers separated by ∼0.01 ly, the required laser power levels are ∼100 W, easily supplied.

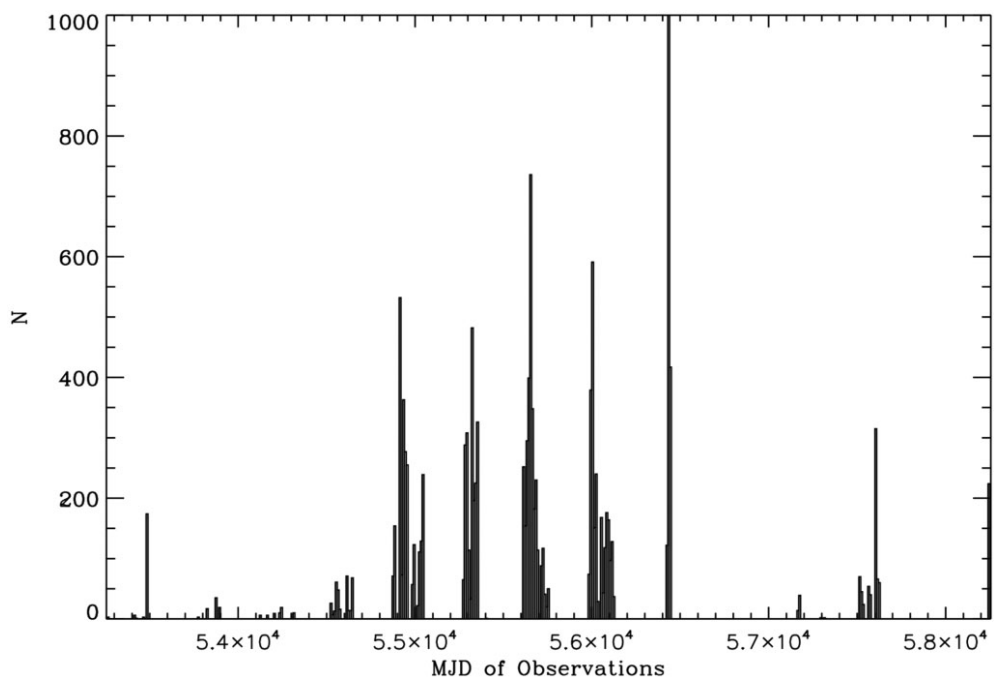

**Figure 7.** The times of observation of the 15 362 spectra obtained between 2004 September and 2018 May, in units of modified julian date. The times of observation span the 14 yr, but the exposure times were 4–20 s, implying large gaps in time between observations.

For reference, continuous-wave lasers on the Earth can operate for many minutes at power levels over 1 MW. The technological challenges of stationing such lasers between the Sun and Alpha Cen seem modest. Short laser pulses with a duration of ∼1 ns from Alpha Centauri would also have been detected if the pulses contained the threshold number of photons during the pulse or within a sequence of pulses. The telescope exposure here of a few seconds simply integrates the arriving photons for the duration of the pulse or their sequence, however long it was. Pulsed and continuous lasers could have been detected, but were not.

The 15 362 spectra sampled only a small fraction of the time during the 14 yr of observations. Fig. 7 shows the times of observation, revealing large gaps. The 15 362 spectra with typical exposure times of 10 s amount to only ∼42 h during 14 yr. The unknown duty cycle of technological lasers located along the line of sight to Alpha Centauri implies that a laser beam may have been missed by the low temporal filling factor of these many observations.

The Alpha Centauri system is an optimal target for continued SETI observations. Technological entities located between the Earth and Alpha Cen could use interferometric telescopes (e.g. Dandumont et al. 2020) to detect signals coming from, and even acquire images of, the technology on Earth, including our radar beacons or city lights (Zuckerman 2002; Gertz 2021; Hippke 2021). Any advanced technological entities at Alpha Cen would know about the technology on Earth, perhaps prompting surveillance and even defence.

An efficient communication network could be constructed with a series of repeater nodes between the Solar system and Alpha Cen (Gertz & Marcy 2022). Such a string of nodes would require less power than a single beam traversing the entire 4 light-year while still offering privacy, 10 GHz bandwidth, and existential security for the engineers. A metre-sized laser located at Alpha Cen, or at nodes along the line of sight, could spotlight a region roughly the size of Earth, allowing us to detect the beam but also allowing avoidance of Earth entirely. Laser communication permits detection or stealth.

The non-detection of optical laser light from Alpha Cen A and B adds to the non-detection of laser light from Proxima Centauri (Marcy 2021), and to the non-detection of laser light from the Solar gravitational lens focal points of both Proxima and Alpha Centauri (Marcy et al. 2021). These add to the non-detections of laser light from over 7000 nearby stars of all masses (Reines & Marcy 2002; Tellis & Marcy 2015, 2017; Tellis et al., private communication). In addition, a $10 \times 14\,\text{deg}^2$ field towards the Galactic Centre exhibited no laser emission (Marcy, Tellis & Wishnow 2022). In addition, searches for broad-band optical and infrared pulses from over 1000 stars have revealed no technological signals (e.g. Stone et al. 2005; Howard et al. 2007; Maire et al. 2019).

The absence of detected optical SETI signals is becoming statistically interpretable as a particular SETI domain within which much of observable parameter space has been surveyed and found empty. This optical SETI desert extends over 10 000 stars within 100 ly. In addition, conventional astrophysics studies of millions of stars within our Milky Way Galaxy have failed to reveal non-astrophysical emission lines. Cannon and Pickering (1922) could have noticed, 100 yr ago, such laser emission in spectra of over 100 000 stars, each examined carefully by eye. Dozens of astrophysics surveys since then would also have found strange emission lines. Indeed, those surveys revealed unexpected emission lines from planetary nebulae, Wolf–Rayet stars, and high-redshift galaxies, to name just three examples. Possible explanations for this absence of optical signals of technology are that any Galactic communication network operates outside optical wavelengths. Alternatively, optical beams may be so narrow and far apart that they rarely intercept the Earth (Forgan 2014). It remains possible that communication is not happening at all among the nearest 10 000 stars, including in the radio domain given the many non-detections (e.g. Tarter 2001, Enriquez et al. 2017; Lebofsky et al. 2019; Price et al. 2020; Margot et al. 2021; Sheikh et al. 2021; Gajjar et al. 2022; Tremblay et al. 2022, Siemion et al. 2014). A looming question is how far the expanding optical SETI desert extends into other wavelengths. A search for monochromatic ultraviolet and infrared transmissions within the Milky Way is warranted.

## ACKNOWLEDGEMENTS

This paper is dedicated to the late Frank Drake who inspired the Search for other civilizations in the Universe and advocated using the world's best telescopes, including radio and optical, to acquire verifiable evidence of their existence. This paper benefitted from valuable communications with John Gertz, Nathaniel Tellis, Beatriz Villarroel, Ben Zuckerman, Franklin Antonio, Brian Hill, Susan Kegley, Carmella Martinez, Paul Horowitz, Andrew Siemion, Roger Bland, and Ed Wishnow.

We are grateful to Dr Pepe and his team for the spectra of Alpha Centauri and thankful for the publicly archived high-resolution spectra obtained with the HARPS spectrometer maintained by the European Southern Observatory (ESO). We also thank the PI and CoIs of the HARPS programme for the design, construction, observations, and reduction of the HARPS spectra, namely, Drs Pepe, Mayor, Benz, Bertaux, Bouchy, Perrier, Queloz, Sivan, Udry, Delfosse, Forveille, Santos, and Moutou.

## DATA AVAILABILITY

This paper is based on data products, reduced spectra, and raw images, obtained with the ESO 3.6-m telescope at the La Silla Observatory. All data are available to the public at the ESO archive: https://www.eso.org/sci/facilities/lasilla/instruments/harps/tools/archive.html and raw data at http://archive.eso.org/eso_archive_main.html.

The data can be retrieved by searching for 'Alpha Centauri', HD 128621, or HD128620.

The spectra were acquired in the programme titled, 'Searching for Earth analogs around nearby stars with HARPS' with PI/CoI Pepe, Lovis, Benz, Bouchy, Mayor, Queloz, Santos, and Udry. The spectra used in this paper were obtained on the ESO 3.6-m telescope and the HARPS spectrometer between 2004 and 2018 within the programme above.

This paper has been typeset from a Microsoft Word file prepared by the author.